%% file: main.tex
\documentclass[a4paper,twoside]{article}

\usepackage{epsfig,subcaption,calc,amssymb,amstext,amsmath,amsthm,multicol}
\usepackage{pslatex,apalike,algorithm2e}
\usepackage[bottom]{footmisc}
\usepackage{graphicx,microtype,placeins,xcolor}
\definecolor{linkblue}{HTML}{3868D9}
\usepackage[unicode,colorlinks=true,linkcolor=linkblue,citecolor=linkblue,urlcolor=linkblue]{hyperref}
\let\mathematicalsup\sup
\usepackage{SCITEPRESS}

\let\sup\mathematicalsup
\hypersetup{
 pdftitle={When Can Quantum Extreme Learning Machines Replace Quantum Reservoirs?},
 pdfauthor={Markus Baumann; Gerhard Stenzel; Jonas Stein; Claudia Linnhoff-Popien},
 pdfkeywords={Quantum Reservoir Computing, Quantum Extreme Learning Machines, Temporal Learning, Fading Memory, Finite Context}
}
\usepackage{etoolbox,orcidlink}
\makeatletter
\patchcmd{\maketitle}{\vskip 0.51in}{\vskip 0.12in}{}{\errmessage{Unexpected title layout}}
\patchcmd{\maketitle}{\vskip -0.07in}{\vskip 0pt}{}{\errmessage{Unexpected title-author layout}}
\renewcommand{\title}[1]{\gdef\@title{#1\\[8pt]}}
\renewcommand{\email}[1]{{\fontsize{9}{11}\selectfont\textit{#1}}\par}
\makeatother
\graphicspath{{figures/}}
\newcommand{\doi}[1]{\href{https://doi.org/#1}{doi:\nolinkurl{#1}}}

\makeatletter\@clubpenalty=\clubpenalty\makeatother

\begin{document}
\title{When Can Quantum Extreme Learning Machines\\Replace Quantum Reservoirs?}
\author{
\authorname{\href{https://qarlab.de/en/our-team/}{Markus Baumann}\,\orcidlink{0009-0007-3575-1006}, \href{https://www.ifi.lmu.de/mvs/de/team/kontaktseite/gerhard-stenzel-911e4c17.html}{Gerhard Stenzel}\,\orcidlink{0009-0009-0280-4911}, \href{https://www.ifi.lmu.de/mvs/en/team/contact-page/jonas-stein-ef892b38.html}{Jonas Stein}\,\orcidlink{0000-0001-5727-9151} and \href{https://www.ifi.lmu.de/mvs/en/team/contact-page/claudia-linnhoff-popien-5eb3fc37.html}{Claudia Linnhoff-Popien}\,\orcidlink{0000-0001-6284-9286}}
\affiliation{\href{https://qarlab.de/en/start/}{QAR-Lab}, Department of Computer Science\\
\href{https://www.lmu.de/en/}{Ludwig-Maximilians-Universit\"at M\"unchen (LMU Munich)}, Munich, Germany}
\email{Correspondence: \href{mailto:markus.baumann@campus.lmu.de}{markus.baumann@campus.lmu.de}}
}
\keywords{Quantum Reservoir Computing, Quantum Extreme Learning Machines, Temporal Learning, Fading Memory, Finite Context.}
\abstract{\looseness=-1\input{sections/abstract}}
\onecolumn\maketitle\normalsize\setcounter{footnote}{0}\vfill

\input{sections/introduction}
\input{sections/models}
\input{sections/background}
\input{sections/dynamics}
\input{sections/evaluation}
\FloatBarrier
\input{sections/conclusion}

\section*{ACKNOWLEDGEMENTS}
This work was supported by the LMU Sustainability Fund (\mbox{EfOiE}), the
German Federal Ministry of Research, Technology and Space (\mbox{BMFTR})
through \mbox{QuCUN}, \mbox{QuaRDS}, \mbox{CAQAO}, \mbox{IgnisQNC}, and
\mbox{Q-NextGrid}, and the Bavarian Ministry of Economic Affairs project
\mbox{6GQT}.
\textit{AI assistance:} OpenAI Codex and \mbox{Anthropic} Claude for ideation,
wording, code, and figures. The authors verified all content and take full
responsibility.

\bibliographystyle{apalike}
{\small\microtypesetup{expansion=false}\bibliography{references}}
\end{document}

%% file: sections/abstract.tex
Quantum reservoir computers (QRCs) retain information from earlier inputs,
whereas reset-window quantum extreme learning machines (QELMs) start afresh
and receive only a recent input window. We establish when this window can
replace the reservoir's memory. Under our stated assumptions and with exact
expectation values, sufficiently expressive QELMs can reproduce reservoir
outputs arbitrarily accurately as their windows grow if and only if the
influence of the omitted past becomes uniformly negligible. However, longer
windows do not always solve the problem. We construct a simple two-qubit
reservoir that performs temporal recall exactly, while every fixed-window
predictor faces the same positive worst-case error over unbounded delays,
regardless of model capacity. Analytical benchmarks and simulations
distinguish missing history from limited features and finite measurements.
The central conclusion is simple: better representations can improve how
available information is used, but cannot recover a past the model never
receives.

%% file: sections/introduction.tex
\section{INTRODUCTION}
\label{sec:introduction}

Whether a quantum extreme learning machine (QELM) can replace a quantum
reservoir computer (QRC) depends on more than model capacity: it depends on
access to the past. We establish when recent inputs contain enough
information to reproduce a reservoir's outputs, how much context is
required, and when persistent memory prevents replacement by any fixed
input window.

QRCs and QELMs offer two ways to process temporal data using quantum features
\cite{fujii2017harnessing,innocenti2023potential}. A QRC carries its quantum
state forward as new inputs arrive, allowing earlier events to influence later
outputs. The reset-window QELMs considered here instead start afresh for each
prediction and receive an explicit window of recent inputs. Their memory
therefore comes from the supplied data rather than from a state retained
between predictions. Time-delayed QELMs demonstrate how this approach can
support time-series prediction without maintaining quantum state throughout
the input sequence \cite{kawanabe2026tdqelm}, as next-generation reservoir
computing does classically \cite{gauthier2021next}. This makes the question of
when a window can replace recurrence practically relevant.

\begin{figure*}[t]
 \centering
 \includegraphics[width=\textwidth]{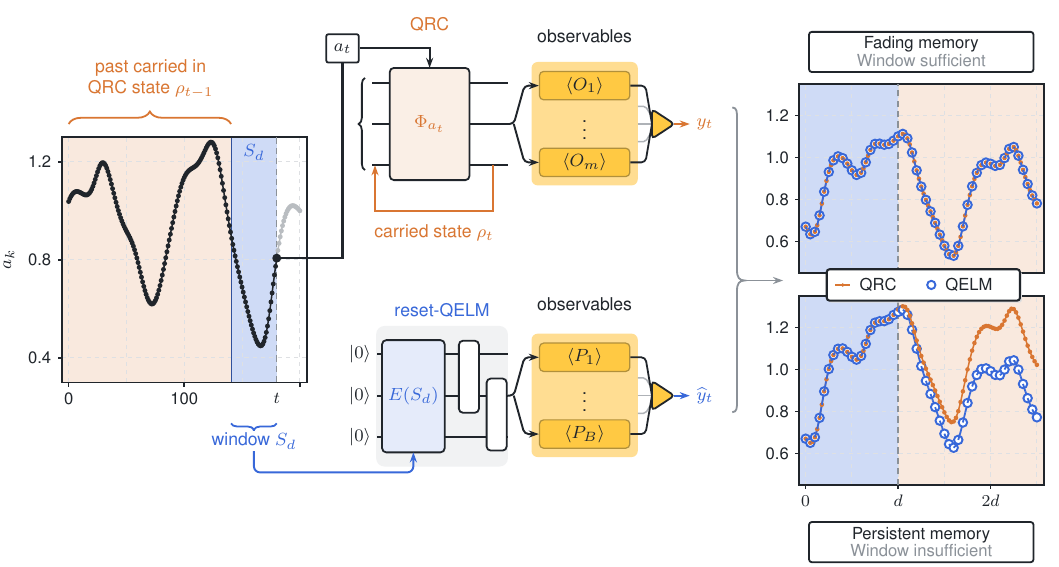}
\caption{Two ways to remember the past. Left: the QRC carries its past forward
 in its quantum state (orange); the reset-window QELM receives only the last
 $d$ inputs (blue). Middle: the QRC updates its carried state with each input;
 the QELM starts from $|0\rangle$ and encodes the window ($E$). Right
 (schematic): an old input leaves the window at the dashed line. An ideal
 window prediction (blue) still matches the QRC (orange) if that input has been
 forgotten (top), but not if its influence persists (bottom).}
 \label{fig:information-interface}
\end{figure*}

The trade-off is between retaining history in a recurrent state and supplying
it explicitly through an input window. A QRC can preserve useful information
from much earlier inputs without replaying them, but its dynamics must retain
that information reliably \cite{baumann2026operating}. A reset-window QELM
avoids carrying quantum state between predictions and makes the available
history explicit, but expanding the window can require additional encoding and
processing resources. Neither is automatically more accurate or more
efficient: the choice depends on how much history the task requires and the
resources available to provide it.

Existing research provides complementary foundations for answering this
question. Reservoir theory studies when dependence on the past
fades \cite{boyd1985fading,martinezpena2023finite}, while quantum
feature-model research establishes which functions sufficiently expressive
models can approximate \cite{goto2021universal}. We connect these
perspectives by separating two requirements: having the historical
information needed for an output, and having a representation capable of
using it.

The distinction is simple. Two input histories may have identical recent
inputs but different reservoir outputs because an earlier event still matters.
A window-only model cannot distinguish these histories, regardless of its
capacity. We quantify the resulting unavoidable error, in the worst case and
in mean square. Under the assumptions of Section~\ref{sec:criterion} and with
exact expectation values, sufficiently expressive QELMs can achieve
arbitrarily accurate emulation as their windows grow if and only if the
omitted past's influence becomes uniformly negligible across the allowed
histories. We also relate the rate of forgetting to the required context
\cite{petreczky2026stability}. The reservoir need not forget everything; only
past information that can still affect the chosen outputs matters.

Longer windows, however, do not always resolve the problem. We construct a
fixed two-qubit reservoir that performs temporal recall exactly using
expectation values. Over unbounded recall delays, every fixed-window
predictor faces the same positive worst-case error, regardless of its
capacity. This separates retained memory from finite context, rather than
quantum from classical computation.

Analytical benchmarks and simulations distinguish missing history from limited
features and finite measurements. Our theoretical guarantees concern what a
model can represent, not whether it can be trained efficiently or implemented
with fewer resources. Together, our results provide a practical starting point
for choosing between recurrent and window-based models: first determine
whether the available context contains the information the task requires; then
decide how best to represent and measure it.

%% file: sections/models.tex
\section{MODELS AND THE EMULATION PROBLEM}
\label{sec:models}

A QRC carries a quantum state forward; a reset-window QELM starts afresh and
receives only a \mbox{window} of recent inputs. We ask when this window
\mbox{contains} enough information to reproduce the reservoir's chosen
outputs.

\subsection{Quantum Reservoirs}

Starting from a specified initial state, a QRC updates the state left by
preceding inputs and produces an output
\cite{fujii2017harnessing,mujal2021opportunities}:
\[
 \rho_t=\Phi_{a_t}(\rho_{t-1}),\qquad y_t=\mathcal C(\rho_t).
\]
Here $a_t$ is the current input, $\rho_t$ is the quantum state, and
$\Phi_{a_t}$ is the input-dependent quantum channel, including any
dissipation. The fixed map $\mathcal C$ collects a finite vector of observable
expectations: the average outcomes of the chosen measurements. A trained
linear readout can be included in $\mathcal C$, with its weights fixed
during emulation.

The target is this output, not the full quantum state. Retained information
matters for our comparison only when it can affect the chosen outputs.

\subsection{Reset-Window QELMs}

A reset-window QELM encodes recent inputs into a freshly prepared state,
applies fixed quantum processing, and measures features. A trained linear
readout combines these features into a prediction
\cite{innocenti2023potential,kawanabe2026tdqelm}.
For a history $h$ ending at step $t$, write $y(h)=y_t$ for the reservoir
output. Its window of the $d$ most recent inputs is
$S_d(h)=(a_{t-d+1},\ldots,a_t)$.
With exact expectation values, the complete
QELM prediction is $\widehat y_t=f(S_d(h))$, where $f$ includes the feature
map and readout. The window length counts input steps, not qubits; quantum
resource requirements depend on the encoding.

The prediction depends only on this window and fixed model parameters.
No additional recurrent quantum or classical state supplies earlier
information. Resetting the quantum register alone would not impose this
restriction if a separate classical memory were retained.

\subsection{The Replacement Question}

Two histories can have identical recent inputs but different reservoir outputs
because an earlier event still matters. A window-only model must give both
histories the same prediction and therefore cannot reproduce both outputs
exactly. Finite measurement sampling does not remove this limitation:
identical windows still produce identical prediction distributions.

Figure~\ref{fig:information-interface} illustrates this distinction
schematically: its blue curves are ideal averages over two equally likely
example histories, not trained QELM predictions, and do not establish
accuracy over all possible histories.

Emulation targets the reservoir's output after the same current input,
not the next value of an external time series. Section~\ref{sec:numerics}
examines forecasting separately.

%% file: sections/background.tex
\section{WHAT A FINITE WINDOW CAN REPRODUCE}
\label{sec:criterion}

A window can support accurate emulation only when histories with identical
recent inputs also produce nearly identical reservoir outputs. We quantify
the uncertainty left by the missing past, then establish when a sufficiently
expressive QELM can approach the best accuracy available from its window.

Consider histories with the same last $d$ inputs but potentially different
earlier inputs. Their largest output difference is
\begin{equation}
 D_d=\sup_{S_d(h)=S_d(h')}\|y(h)-y(h')\|.
 \label{eq:diameter}
\end{equation}
The supremum runs over all allowed histories. The norm is fixed and reduces
to the absolute value for a scalar output. Thus $D_d$ measures how much the
output can vary when everything the window model receives is held fixed.

For a scalar output, each window is compatible with a range of reservoir
outputs. Predicting its midpoint minimizes the largest possible error:
any other prediction moves closer to one end but farther from the other.
If the range is $0.2$ to $0.8$, for example, predicting $0.5$ gives the
smallest worst-case error, $0.3$. Across all windows,
\begin{equation}
 \inf_f\sup_h |y(h)-f(S_d(h))|=\frac{D_d}{2},
 \label{eq:window-law}
\end{equation}
where $f$ ranges over all functions of the window. This limit applies
regardless of architecture or capacity: it uses no quantum structure and
binds every predictor that sees only the last $d$ inputs. A longer window
can reduce it by
distinguishing histories that a shorter window treats as identical. If
$D_d\to0$, arbitrarily small errors become possible in principle. If $D_d$
does not vanish, additional capacity cannot remove the error floor.

Approaching this limit with a QELM requires suitable dynamics and model
capacity. Assume a finite-dimensional reservoir, a compact input alphabet,
input-continuous quantum channels, and fixed observable outputs. Every
finite input continuation must be allowed after every admissible earlier
history. These conditions make the midpoint prediction continuous in
the window.

For each window length, let $\mathcal Q_d$ be a reset-window QELM family that can
approximate every continuous window function uniformly as capacity grows,
using exact expectation values. Then, for scalar outputs,
\begin{equation}
 \inf_{f\in\mathcal Q_d}\sup_h
 |y(h)-f(S_d(h))|=\frac{D_d}{2}.
 \label{eq:qrc-qelm-equivalence}
\end{equation}
The family can approach the window's information limit arbitrarily closely.
This is an infimum: a particular finite model need not attain it exactly.

Consequently, under these assumptions, arbitrarily accurate QELM emulation is
possible precisely when $D_d\to0$ as the window grows. The omitted past must
become negligible uniformly across all allowed histories, not only for typical
test examples. The same criterion holds for finite vector outputs, although
their unrestricted fixed-window optimum lies between $D_d/2$ and $D_d$.
Randomization cannot lower the scalar bound on worst-case
expected absolute error, because histories with the same window receive the
same prediction distribution.

The limit is not an artifact of worst-case analysis. For a scalar output and
histories drawn from any probability distribution, the best window predictor
in mean square
is the \mbox{conditional} mean of $y$ given the window, and its error is the
remaining variance:
\begin{equation}
 \inf_f\mathbb E\,|y(h)-f(S_d(h))|^2
 =\mathbb E\bigl[\operatorname{Var}(y\mid S_d)\bigr]\leq\frac{D_d^2}{4}.
 \label{eq:mean-square-law}
\end{equation}
The inequality holds because all outputs compatible with one window lie in
an interval of length at most $D_d$. It is tight: two equally likely
histories that share a
window and differ in output by $g$ force a root-mean-square error of at
least $g/2$. Hence $D_d/2$ is also the supremum of the root-mean-square
floor over distributions of histories. The floor in
Eq.~\eqref{eq:mean-square-law} also binds randomized predictors such as
finite-shot readouts, and $\mathcal Q_d$ approaches it under the
assumptions above, since continuous window functions are dense in mean
square.

For inputs in $[0,1]$, repeated encoding and Bernstein approximation give an
explicit growing QELM construction
\cite{goto2021universal,yu2024nonasymptotic}. Its resource costs are discussed
in Section~\ref{sec:finite-resources}. This existence result neither
guarantees efficient training nor makes the fixed architectures in
Section~\ref{sec:numerics} universal. The replacement question is now one of
memory: when do older inputs lose their influence, and how much context makes
their omission acceptable?

%% file: sections/dynamics.tex
\section{MEMORY DECAY AND THE COST OF REPLACEMENT}
\label{sec:dynamics}

Accurate emulation does not require the reservoir to lose all information
about its past. Only the influence of inputs outside the window on the
chosen outputs must become negligible. How quickly this influence decays
determines the required context; persistent influence can prevent
finite-window replacement.

\subsection{When Longer Windows Are Enough}
\label{sec:observable-forgetting}

A simple way for older inputs to become irrelevant is for every update to
bring different states closer together \cite{chen2020temporal}. Suppose
each input-dependent channel contracts trace distance by the same factor
$0\leq\eta<1$:
\begin{equation}
 D_{\rm tr}(\Phi_a(\rho),\Phi_a(\sigma))
 \leq\eta D_{\rm tr}(\rho,\sigma).
 \label{eq:incremental-channel}
\end{equation}
Here $D_{\rm tr}(\rho,\sigma)=\tfrac12\|\rho-\sigma\|_1$ is the trace
distance. After $d$ shared inputs, a scalar output
$y=\operatorname{Tr}(O\rho)$ satisfies
\begin{equation}
 D_d\leq 2\|O\|_\infty\eta^d,
 \label{eq:contraction-bound}
\end{equation}
where $\|O\|_\infty$ is the largest absolute eigenvalue of the output
observable. The unavoidable window error therefore decreases exponentially,
so a sufficiently long window meets any fixed positive error tolerance.

Full-state contraction is stronger than necessary. Some state differences
may persist without ever affecting the chosen outputs. Others may be
invisible now but become visible after later inputs. Only differences
invisible under every possible continuation can safely be ignored.

We therefore keep state differences that input histories can create and
remove those that no future output can distinguish. This is a
reachable--observable reduction \cite{grigoryeva2021canonicalization}. Let $A_a$ describe the
effect of input $a$ on the remaining differences. Their slowest asymptotic
decay is characterized by the joint spectral radius
\cite{rota1960joint,jungers2009joint},
\begin{equation}
 r_{\rm vis}=\lim_{d\to\infty}
 \sup_{a_1,\ldots,a_d}\|A_{a_d}\cdots A_{a_1}\|^{1/d}.
 \label{eq:visible-radius-definition}
\end{equation}
This quantity considers complete input sequences rather than isolated
updates; its value does not depend on the norm. Under the reservoir and input
assumptions of Section~\ref{sec:criterion},
\begin{equation}
 D_d\to0\quad\Longleftrightarrow\quad r_{\rm vis}<1,
 \qquad\lim_{d\to\infty}D_d^{1/d}=r_{\rm vis}.
 \label{eq:visible-memory-theorem}
\end{equation}
Thus, arbitrarily accurate emulation is possible exactly when all reachable
differences that can affect future outputs decay uniformly. Under
full-state contraction, $r_{\rm vis}\leq\eta<1$. This criterion specializes
\cite[Theorem~III.1]{petreczky2026stability}.

\begin{figure*}[t]
 \includegraphics[width=\textwidth]{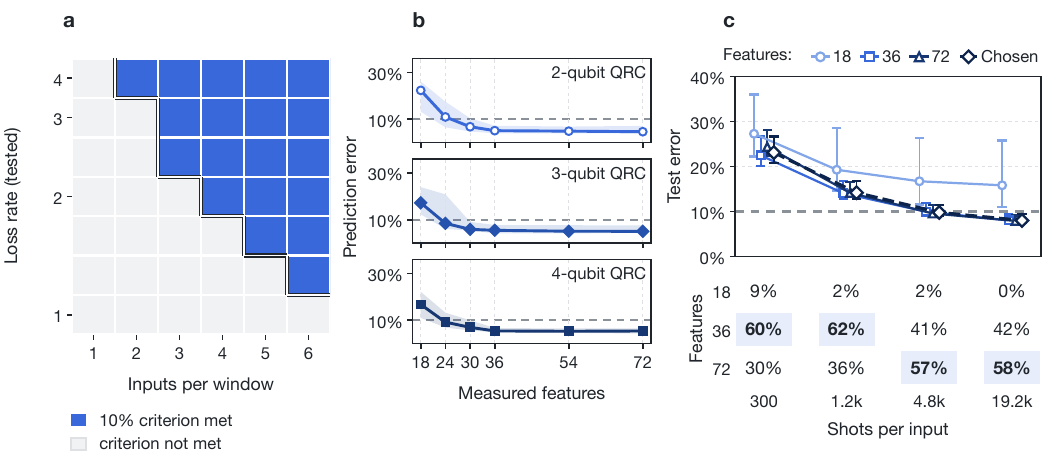}
 \caption{Three resources, three questions.
 (a) Enough history? Blue: for all three reservoir sizes, the sampled influence
 of inputs older than the window stays within the 10\% tolerance. Faster forgetting
 (higher loss) needs shorter windows; at loss rate 1 or lower, no tested window
 suffices.
 (b) Enough features? Error falls with more measured features, with little gain
 beyond 36 (dashed: 10\%).
 (c) Enough shots? More shots lower the error and let validation pick more
 features (table: share of picks; bold: most frequent).
 Medians with 10th--90th percentiles across realizations; separate studies per
 panel (Section~\ref{sec:numerics}).}
 \label{fig:resource-diagnosis}
\end{figure*}

The decay rate determines how much history is needed. For a scalar output
and absolute-error tolerance $\varepsilon>0$, the shortest sufficient
window is
\begin{equation}
 d_\varepsilon=\min\{d:D_d<2\varepsilon\},
 \label{eq:context-length}
\end{equation}
with $d_\varepsilon=\infty$ if no finite window suffices. This window permits
worst-case error below $\varepsilon$, both for an unrestricted predictor
and, under the approximation assumptions of Section~\ref{sec:criterion}, for a
sufficiently expressive QELM. When $0<r_{\rm vis}<1$,
\begin{equation}
 \lim_{\varepsilon\downarrow0}
 \frac{d_\varepsilon}{\log(1/\varepsilon)}
 =\frac{1}{-\log r_{\rm vis}}.
 \label{eq:window-complexity}
\end{equation}
The required window thus grows logarithmically with the inverse error
tolerance. Slow
forgetting does not itself prevent replacement, but requires longer context.
This describes the information requirement and leaves learning and
implementation costs aside.

\subsection{When Finite Windows Cannot Suffice}

Longer windows cannot resolve every memory requirement. We construct one
fixed two-qubit reservoir in which an old input leaves an observable trace
that survives arbitrarily many later inputs. The same reservoir provides
a counterexample for every finite window length.

Each input consists of a drive value $u\in[0,1]$ and a binary flag $q$
controlling a local pulse on the first qubit. Starting from the same initial
state, two histories differ only in whether one old input triggers this pulse.
The pulse changes the antisymmetric singlet population. Both histories then
receive identical inputs with $q=0$, so no further pulses occur, and the
subsequent symmetric evolution with collective loss preserves the difference
\cite{zanardi1997noiseless}. The input rule is unchanged over time.

Ordinary two-qubit correlations reveal the stored information. The singlet
state $|\psi^-\rangle=(|01\rangle-|10\rangle)/\sqrt2$ has projector
\begin{equation}
 P_-=|\psi^-\rangle\!\langle\psi^-|
 =\frac{II-XX-YY-ZZ}{4},
 \label{eq:protected-singlet}
\end{equation}
where $I$ is the identity and $X,Y,Z$ are Pauli operators, with
$XX=X\otimes X$ and similarly for the other terms. Its population
$y=\operatorname{Tr}(P_-\rho)$ is recovered from these three correlations.

Let the pulse create a population difference $\delta>0$. Arbitrarily long
shared, unpulsed continuations preserve it, giving
\begin{equation}
 D_d\geq\delta\qquad\text{for every finite }d.
 \label{eq:protected-mode}
\end{equation}
More generally, any reachable output gap preserved by freely repeatable
common inputs gives this obstruction; conservation need not hold for every
allowed input. Equation~\eqref{eq:window-law} forces a worst-case error of
at least $\delta/2$ for every stateless window-only predictor, quantum or
classical. Here the preserved gap also gives $r_{\rm vis}=1$.

The recall task is to report, after an arbitrarily long delay, the
population ($0$ or $\delta$) set by the old flag. Exact QRC expectations realize
this target. Once the flag leaves the window, a reset-window QELM receives
identical inputs for the two targets and cannot reproduce both. Neither
additional capacity nor any fixed enlargement of the window eliminates
the worst-case gap over unbounded delays. Nor is this only a worst case:
for a flag set with probability one half, Eq.~\eqref{eq:mean-square-law}
leaves every window predictor a root-mean-square error of at least
$\delta/2$ once the flag has left the window, against zero for the QRC.

This separates retained memory from finite context, not quantum from classical
computation. It does not rule out a window covering a bounded task horizon.
The protection rests on the symmetry of the collective loss: local loss
reduces the gap, while asymmetric collective loss breaks exact protection
without necessarily erasing memory quickly.

\subsection{From Sufficient History to Accurate Predictions}
\label{sec:finite-resources}

Enough history does not guarantee an accurate representation or precise
measurements. An explicit construction separates these requirements.
For scalar inputs in $[0,1]$ and a scalar output,
we assume uniform state contraction and
Lipschitz input dependence: small input changes have proportionally bounded
effects on the updated state. These
conditions are stronger than observable forgetting alone.

Each input is encoded into $r$ separately prepared copies, and a prescribed
readout approximates the midpoint prediction of Section~\ref{sec:criterion}.
More copies \mbox{provide} richer functions of the same window, not older \mbox{history}.
Averaging $Q$ independent measurements of this bounded readout gives, by
Hoeffding's inequality \cite{hoeffding1963probability},
\[
 \begin{aligned}
 &|y(h)-\widehat f_r(S_d(h))|\\
 &\quad\leq\underbrace{D_d/2}_{\text{missing history}}
 +\underbrace{\varepsilon_{\rm rep}}_{\text{representation}}
 +\underbrace{\varepsilon_{\rm shot}}_{\text{measurements}}.
 \end{aligned}
\]
For each fixed history, but not simultaneously over all histories, this
holds with probability at least $1-\zeta$, where $0<\zeta<1$ is the allowed
failure probability. With exact expectations, the sampling term disappears
and the remaining bound is uniform. At fixed window and confidence, the
representation and sampling bounds decrease as $1/\sqrt r$ and
$1/\sqrt Q$, respectively.

One sufficient implementation uses $rd$ qubits and a general table of
$(r+1)^d$ readout coefficients. Computing them requires reachable-output
extrema and is not included in the resource counts. This constructive upper
bound is neither an efficient learning algorithm nor an exact decomposition of
fitted-model error.

%% file: sections/evaluation.tex
\section{NUMERICAL EXPERIMENTS}
\label{sec:numerics}

Our experiments distinguish three obstacles to replacement: missing history,
limited use of available information, and noisy measurements. We examine
these separately, test whether older inputs help on a prediction task, and
use analytical benchmarks to check the theoretical error floor.

The main studies compare a six-qubit QELM with two-, three-, and four-qubit
reservoirs. Reservoirs evolve under random $XX$ couplings, a transverse field,
an input drive, and local loss; higher loss means faster forgetting. The QELM
encodes one input per qubit, evolves without loss, and measures single-qubit
Pauli features; ridge penalties are selected on validation data. These fixed
models are not the growing universal family of Section~\ref{sec:criterion}.
The simulation studies use 12 independent realizations; the separate
prediction study uses 24. Training, validation, test, and diagnostic data are
separate. Normalized root mean squared error (NRMSE) is expressed as a
percentage of the target standard deviation. A history or emulation setting
meets the 10\% tolerance when at least 10 of 12 realizations pass. Code:
\url{https://github.com/eybmits/qrc_vs_qelm}.

\begin{figure*}[t]
 \centering
 \includegraphics[width=\textwidth]{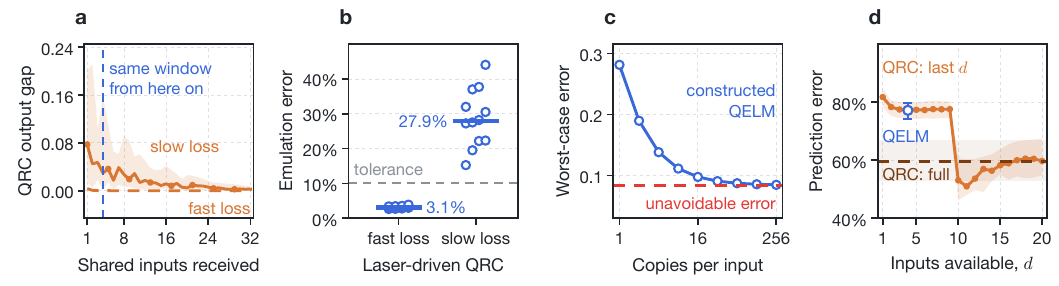}
 \caption{From missing history to error.
 (a) Under slow loss, QRC outputs from two different pasts stay apart long
 after the four-input windows coincide (blue line); under fast loss they agree
 at once.
 (b) Accordingly, the four-input QELM emulates the laser-driven QRC within the
 10\% tolerance only under fast loss (circles: 12 realizations).
 (c) Capacity cannot repair this: in a separate benchmark, a constructed QELM
 approaches, but cannot pass, the unavoidable window error (red).
 (d) On NARMA10, the QRC restricted to its last $d$ inputs regains its
 full-history accuracy (dashed: a reference, not a floor) only for $d\geq10$;
 the four-input QELM matches
 the restricted QRC.
 Lines and bars: medians; bands and whisker: 10th--90th percentiles.}
 \label{fig:long-context}
\end{figure*}

\subsection{When Is the History Sufficient?}

We give different pasts exactly the same continuation. The remaining variation
of the outputs around their mean, the sample counterpart of the conditional
variance in Eq.~\eqref{eq:mean-square-law}, reveals the influence of earlier
inputs. The distance from that mean to the QELM prediction is the residual; it
combines representation, fitting, and estimation effects. The two squared
terms add exactly, but neither estimates the theory's worst-case error or
asymptotic forgetting rate.

Faster forgetting reduces the required window
[Figure~\ref{fig:resource-diagnosis}(a)]. At the strongest tested loss, two
inputs meet the history criterion across all three sizes. At weak loss, no
tested window of up to six inputs does; longer windows remain unresolved. In
all 27 tolerance/replication sensitivity variants, the required context
remains non-increasing in the loss rate. In a separate five-qubit study,
extending the window from two to ten inputs reduces history error in every
realization at both tested losses, but sufficient history does not always
yield accurate fitted QELMs.

Measured Santa Fe laser inputs~\cite{weigend1993timeseries} illustrate the
same distinction. Under identical subsequent inputs, two different pasts
remain distinguishable long after the four-input windows coincide when loss
is slow, but not when it is fast [Figure~\ref{fig:long-context}(a)].
Accordingly, the QELM's median error in emulating the laser-driven
reservoir's output (not the laser signal) is 3.1\% with fast forgetting and
27.9\% with slow forgetting [Figure~\ref{fig:long-context}(b)].

\subsection{Using the Available Information}

Enough history does not ensure accurate prediction. With fast forgetting and
six recent inputs, the 18-feature QELM usually misses the tolerance, whereas
the same reservoir restarted from that window succeeds. Increasing to 36
features meets the tolerance in every realization across all three sizes
[Figure~\ref{fig:resource-diagnosis}(b)]. This diagnoses the tested features
and fitting procedure and is not an impossibility result for QELMs. Fitted
accuracy also need not improve monotonically with window length. Matched
classical quadratic delay controls outperform the tested QELMs in the
fast-forgetting comparisons.

Measurement precision introduces a further limitation. Increasing
shots from 300 to 19,200 per input reduces median test error from
23.1\% to 8.0\%, improving every realization in the exploratory shot study
[Figure~\ref{fig:resource-diagnosis}(c)]. Validation generally selects fewer
features with scarce measurements and more with precise measurements.
Extra features are useful only if estimated reliably
\cite{hu2023sampling,ahmed2025finite}. \mbox{Reservoir} targets remain exact,
isolating QELM sampling noise; hardware errors and measurement backaction
are excluded.

\subsection{Does Older History Help Prediction?}

Synthetic NARMA10 tests prediction rather than emulation: both models learn
to predict the next target value from the input
sequence~\cite{fujii2017harnessing}.
Across 24 realizations, median test error is 59.7\% for the recurrent
QRC and 77.2\% for a four-input QELM. Restricting the same QRC to four inputs
raises its error to 77.3\%. This control restarts the reservoir for each
prediction, processes only the window, and refits the readout. The recurrent
QRC outperforms both four-input models in every realization.

Longer windows substantially improve the restarted reservoir
[Figure~\ref{fig:long-context}(d)]: older inputs help.
Because features and readouts change with the window length, a single causal
delay cannot be isolated; longer replay also requires more evolution. A
twelve-input classical quadratic predictor achieves 10.3\% error. The finding
therefore concerns insufficient context and shows neither quantum necessity
nor the failure of every finite window.

\subsection{Reaching the Theoretical Limit}
\label{sec:sharpness}

Separate analytical benchmarks test attainability. A one-qubit reservoir
retains 80\% of its previous output at each step and combines it with a
response to the new input. For an eight-input window, its optimal worst-case
absolute error is $0.8^8/2=0.08388608$.

With a quadratic input response, two encoding copies per input represent the
midpoint prediction exactly through an analytically specified readout. A
\mbox{16-qubit} statevector calculation confirms this readout on tested
windows to rounding precision.

With an exponential response, a prescribed construction approaches the same
limit. Increasing from one to 256 copies per input reduces \mbox{worst-case}
\mbox{absolute} error from $0.281322$ to $0.084753$
[Figure~\ref{fig:long-context}(c)]. The large-copy results use factorized
exact expectations, not full-register simulations; better readouts at fixed
capacity may exist.

These constructed readouts are not trained models; they show that
capacity can reduce representation error while unavailable history leaves
an unavoidable error floor.

%% file: sections/conclusion.tex
\section{DISCUSSION AND CONCLUSIONS}
\label{sec:summary}

Reset-window QELMs can replace QRCs when recent inputs provide enough
information to reproduce the required outputs. Under our stated assumptions
and with exact expectation values, sufficiently expressive QELMs can achieve
arbitrarily accurate emulation as their windows grow if and only if the
influence of the omitted past vanishes uniformly. Fast forgetting permits
shorter windows; slower forgetting may require more context.

Replacement can fail when earlier events continue to affect the outputs. Our
two-qubit recall example makes this limitation explicit: the reservoir retains
information about an old input over arbitrarily long delays, whereas every
fixed-window predictor faces an unavoidable worst-case error. Increasing model
capacity cannot remove this gap, and choosing a larger but still fixed window
does not eliminate it over \mbox{unbounded} delays. Because the limit is
\mbox{independent} of architecture and also holds in mean square, it applies
to any finite-context model of a reservoir, quantum or classical.

Analytical benchmarks and simulations distinguish this information limit from
errors caused by limited features and finite measurements. Having enough
history is thus only the first requirement for accurate emulation. The
theoretical guarantees establish what models can represent; efficient training
and cheaper implementation are separate questions. Hardware errors,
measurement backaction, and constrained input sequences are left for future
work.

The practical conclusion is to assess memory before capacity: a sufficiently
expressive QELM can replace the reservoir when recent context suffices. When
the required history cannot be supplied through the window, more features
cannot make up for it; retained memory is needed.